# GENEALOGY TREE: UNDERSTANDING ACADEMIC LINEAGE OF AUTHORS VIA ALGORITHMIC AND VISUAL ANALYSIS

Sandra Anil[1], Abu Kurian[1], Sudeepa Roy Dey[1], Snehanshu Saha[1], Ankit Sinha[1]

[1]*Department of Computer Science and Engineering, PESIT Bangalore South Campus, Bangalore*

Correspondence: sudeepar@pes.edu

**ABSTRACT**

Ancestry and Genealogy tree are proven tools to determine the lineage of any person and establish dependencies among individuals. Genealogy tree can be exploited further to gain information about the researcher and his scholastic lineage which is of paramount importance in today's world of computer technology. This insight into academic genealogy could be ways of helping PhD students achieve academic socialization within the discipline, by making explicit connections that may be influential. Awareness of his scientific heritage, gives the user a broader perspective of his own research project. This paper also highlights and investigates how this academic network is exploited by certain researchers using various visualization tools. It was observed during this work that the credibility and influence factor is determined by the various citations obtained by an author and to improve their rankings in various forums, they tend to collaborate in their academic circle and boost their citation count. A recent trend among researchers is to form communities based on their academic relationships and rely on copious citations for their mutual benefit. Tracing the genealogical relationships can be helpful in detecting such communities and also create a more quality aware metrics using a lineage independent model for computation of author level metrics.

## INTRODUCTION

Genealogy is an account of descent of a person, family or group from an ancestor or from older forms. It is the study of the history of the past and present members of a family or families. Historical records are used for genealogical research. Over the years, the number of people pursuing PhD has increased, leading to an exponential growth of the academic genealogical tree. With this rising number, keeping track and documentation of scholastic relationships between scientists has become difficult. An attempt in this direction has been made by the American Mathematical Society, by means of their Mathematics Genealogy Project. Their objective is to catalogue the complete mathematics community. It gives information of an author, his ancestry and lineage in the tree, along with his dissertation and year of being awarded the degree. A similar approach has been put forth in this paper for the discipline of Computer Science along with some new metrics. Mining huge databases is a complex task and lot of algorithms are derived to create a database of authors [1]. The genealogy tree will hold information about all the scientists who have contributed to the field at various research-level.

A database is built from the contributions of the scientists who give inputs like their supervisors, dissertation details, year and affiliating institute. A graph database is formed which is then searched based on user query. The tree obtained can be further based on two criteria: author or domain. The tree based on author describes the author's heritage and his descendants. Details about the author's degree are also provided in this genealogical tree.

Since Computer Science can be perceived as an umbrella housing many domains which has multiple research areas within them, the domain based tree traces the complete hierarchy of scientists who have contributed to it significantly.

The proposed model will be used further to trace various citation patterns among authors under the same advisor i.e. siblings to each other, who form communities to inflate their citation count by mutually citing one another consistently. Such networks are also observed within an advisor and advisee as well. The model proposes a threshold, which is the ratio of community by total citation and the authors who exceed this threshold are identified

*The scibase and scientometric modeling effort is endorsed and supported by IEEE Computer Society Bangalore Chapter.*

as potential lineage or community dependant authors. Our motive is to trace and highlight such communities and their patterns of citations.

## MOTIVATION AND TECHNICAL CONTRIBUTION

With the exponential growth in researchers and their publication in various journals, the need to trace the quality of work and rank the emerging authors has also increased [3]. A genealogy tree will not only help explore the pedigree and the ancestry of an author and the domain, it will also be used to investigate the citations received and do the in-depth analysis. With this motivation, a similar software model has been put forth by this paper for the department of Computer Science.

The project is based on a graph database which is built by contributions of the scientists who provide input details like dissertation, and year and institute of procuring degree. A genealogy tree is created with nodes as authors/supervisors and link represents advisees. Finally, a revenue model is also presented where an Author Lineage Score is calculated based on various parameters and the authors are classified regarding this.

## COMMUNITY CITATION

Community of authors is defined as a group of authors who collaborate with each other and frequently cite each other's work to increase their citations count. These communities can be formed as a network of an advisor and his students or alternately between students under the same advisor in which case it will be a sibling network.

The proposed method is to subdivide the author citation matrix into block matrices for each individual author. A block matrix is a collection of sub matrices where each sub matrix is an independent matrix having rows and columns. Each block matrix represents the local network of an author at each level. Block matrix is an adjacency matrix i.e. if A[i, j] = 1, then j is a member of local network of i. The first row of the matrix represents the children of the author i.e. one level below the author in the genealogical tree. Similarly, the second row represents the children of the author two levels down, and the third and fourth row represent the parent and grandparent of the author respectively.

Fig 1. **Block Matrix division.**

The block matrix division in fig 1. shows A, B, C and D are the sub matrices. Each sub matrix is a dedicated adjacency matrix of an author. In the sub matrix A, author $A_1$ is one hop below A in the genealogical tree. $A_2$ is two hops below A, $A_4$ is two hops above A in the genealogical tree. Similarly, the local network of authors B, C and D are also represented in the block matrix as independent sub matrices.

Determining the local network of an author using block matrices reduces the computational complexity, since traversal of the JSON file is eliminated. The time complexity of the algorithm, to determine the local network by parsing the JSON file is $O(n^3)$. Block matrix division gives a time complexity of $O(n)$.

Community Citation is defined as the total citations obtained by the author from his citation community. It is a subset of the Genealogical Citations of the author.

A special case of community citation is the copious citation. Copious citation refers to citations between pair of authors. The authors mutually cite each other's work and by this they both benefit from it.

A threshold value for detecting a community must be determined from historical data. An appropriate trend detection algorithm must be used to decide the threshold. The threshold will have a lower bound and upper bound i.e. the threshold will be a finite range in practice.

The total genealogical citations are obtained by adding the citations of the authors present in the local network, from the author citation matrix. The ratio of the genealogical citations with the total citation is computed and compared against the threshold and the author is labelled as lineage dependant if the ratio is greater than the threshold.

## NON-GENEALOGICAL CITATION

Non-genealogy citations (NGC) of an author are defined as citations obtained from authors who are not present in the genealogy network of that author. Non-genealogy citations can be computed as the difference between total citations and genealogy citations.

$$\begin{array}{c c} & \begin{array}{cccc} A & B & C & D \end{array} \\ \begin{array}{c} A \\ B \\ C \\ D \end{array} & \left[ \begin{array}{cccc} 10 & 7 & 0 & 15 \\ 21 & 19 & 0 & 1 \\ 15 & 0 & 3 & 12 \\ 0 & 17 & 0 & 1 \end{array} \right] \end{array}$$

Fig 2. **All Author Matrix.** In the matrix, for author A self-citations are 10. Author A has cited Author B 7 times, Author D 15 times and has not cited Author C even once. The complete matrix can be interpreted similarly.

The citations of authors are represented in an all author matrix. Each element A[i,j] denotes the number of times author j has cited author i. Every author is assigned a unique id in the all author matrix.

The following cases may arise while computing the NGC from an author network:

1. Unique Name Case: The authors name is unique in the complete network.
2. Multiple Name Case: There is more than one author having the same name.
3. Two Advisor Case: The author will have more than one advisor, irrespective of the uniqueness of his name.
4. Multiple Name and Two Advisor Case: Authors sharing the same name, few of them will have multiple advisors.

The total citations of an author represented by Y are obtained from the all author matrix. To compute the genealogical citations first the authors who are part of the genealogy network of the author are identified.
Then the sum of citations of all these authors, represented by X, is obtained from all author matrices. NGC is calculated as

$$NGC = Y - X$$

## TECHNICAL IMPLEMENTATION

*1. Creation of Database*
A sample graph database was created to test the Community Detection method. Each node in the database had 8 fields. These fields were name of the author, the author's teacher (labelled as Level 1), the author's students (labelled as Level 2), the author's Ph.D. thesis, the Institute at which he acquired his Ph.D., the Country of origin, domain, and the number of overall citations he has received. Each teacher or student present in the above-mentioned attribute Level 1 or Level 2 will be stored as a key-value pair. The data structure will store the name along with the number of citations to the author under consideration. This will help to detect any irregular or false citations that the author under study may be receiving by his local community.

*2. The Visualization toolbox*
A web application was created to visualize community detection. The website [8] offers 3 options. First is to display the local network of any given author, the second option is to display the details of the author and the third option is for community detection. When the user refers to the local network it is the community around the author that includes the author his teacher, his teacher's teacher, the author's students and the student's student. When an author is queried and the third option is selected the software will check whether there is a presence of local citations within the author's local community and display the results.

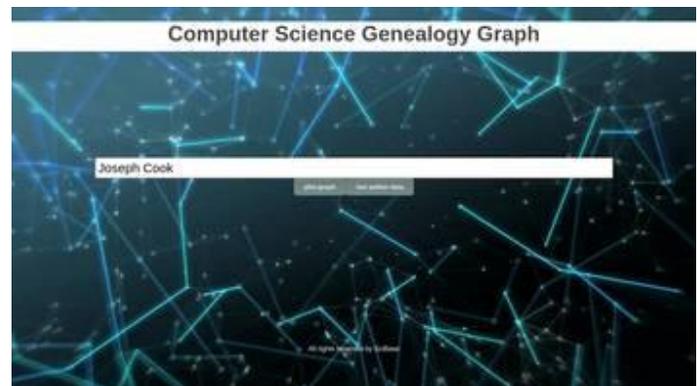

Fig. 3. **Home Page of the website. Search an author page**

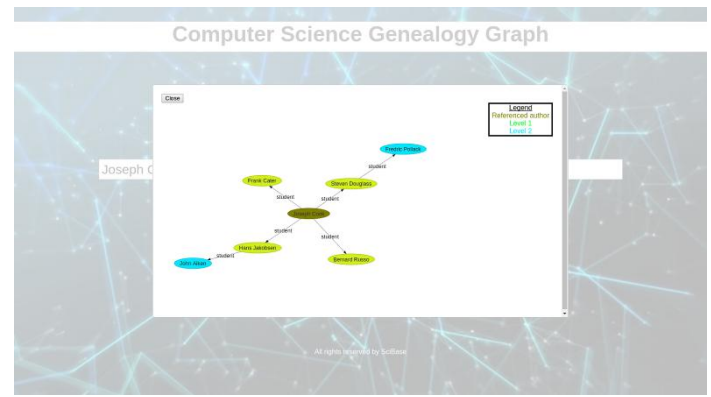

Fig. 4. **Genealogy tree of the author.**

*The scibase and scientometric modeling effort is endorsed and supported by IEEE Computer Society Bangalore Chapter.*

Fig 3 and 4 depict the pages where an author Joseph cook is search and fig 4. Genealogy tree of Joseph Cook is shown. The authors represented by green nodes are the nodes one level below, Joseph Cook i.e his advisees and the authors represented by blue nodes are the nodes two levels below i.e advisees of his advisees.

The visualization shown can be viewed and interacted with, by visiting the website http://gt.sahascibase.org/

For the creation of a graphical view, a JavaScript based library, vis.js, was used. When the user clicks the network display button, a GET request is sent to our backend server which in turn requests the Neo4J database with recursive queries.

Initially, 2 empty arrays NODES and EDGES are defined and the name of author for whom the graph is being generated, is pushed in the NODES array.

Each recursive query then returns the next level of related authors in the community. [9] For each query result, the name of all the next level authors is pushed in the NODES array, and a JavaScript object in the form
{
  'from' : currentAuthor,
  'to' : nextLevelAuthor
}
is created and pushed in the EDGES array. At the end of recursive query calls, the formed NODES and EDGES arrays are sent back to the frontend, where with a call to vis.js library function along with providing it the NODES and EDGES array in proper format, a graph as shown in Figure 4 is created.

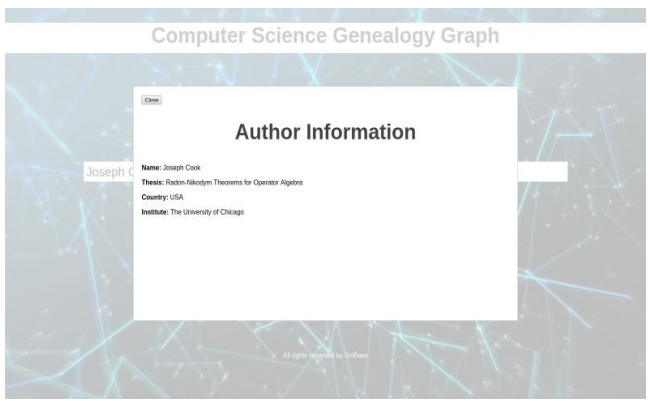

Fig. 5.  **Author information.**
The profile of Joseph Cook i.e institute, thesis and country are displayed

3. *Final Neo4j database*

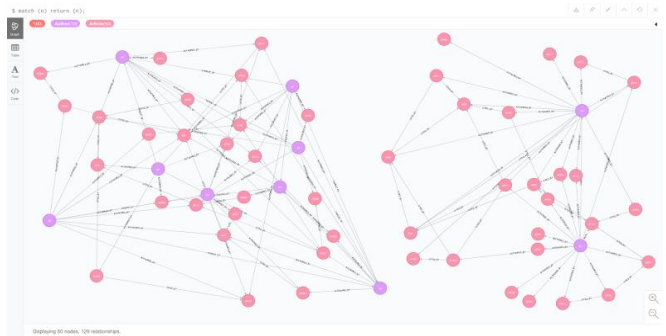

Fig. 6. **Neo4j database.**

There are two types of nodes in the database as shown in fig 6.. The purple nodes are the author nodes and the pink nodes are the article nodes. There are three types of relations PARENT_OF, CITED_BY and AUTHORED_BY. The author nodes are connected by the PARENT_OF relation. The author node which is connected to another author node by this relation is the advisor of that author node. The articles are connected by the CITED_BY relation, indicating articles which have been cited by other articles. The author and article nodes are connected by AUTHORED_BY relation indicating the author of each article

## CONCLUSION AND FUTURE WORK

This paper has highlighted certain key aspects of the author metrics and investigated the citations based on the source and not just the count. The genealogy tree structure is exploited to generate the adjacency matrix, which is then used to derive various metrics such as Genealogy citations (GC), Non-genealogy citations (NGC) which have potential to filter all the contribution towards unethical citation boosting. The authors of this paper humbly put across the following contributions:

- A sample database of authors and fields were name of the author, the author's teacher (labeled as Level 1), the author's students (labeled as Level 2), and the author's Ph.D. thesis, the Institute at which he acquired his Ph.D., the Country of origin, domain, and the number of overall citations he has received.
- Author level metrics derived exclusively with genealogy tree as the backbone. These metrics are further validated through the web application.
- Community detection Algorithms designed to exploit the genealogy tree to the optimum and trace the susceptible population that influence the citation counts of an author in various ways.

*The scibase and scientometric modeling effort is endorsed and supported by IEEE Computer Society Bangalore Chapter.*

- A web application is created to visualize community detection as well.
- Devise a scoring model to highlight an author's influence across his /her lineage and community.
- The visualization shown can be viewed and interacted with, by visiting the website http://gt.sahascibase.org/

Investigating the pattern of citations and measuring the true impact of a researcher has always been a debatable yet important aspect of scientometrics. With this software, the authors hope to create a justified visualization model and put forth new author level metrics that can quantify the influence of an author in terms of pure quality research work, thereby converging quantity and quality of research. The software can be used for further analysis of metrics like sibling citations and recursive copious citations. In future, this model can serve as a visual tool to calculate and perform in-depth analysis of citation network.

**BIOGRAPHY**

**Sandra Anil** is a final year B.E student and is a active member of scibase project. She has interest in graph databases and machine learning.

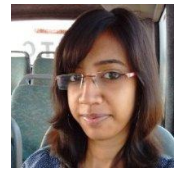

**Abu Kurian** is currently pursuing his bachelors in Computer Science and engineering. An active member of scibase project familiar with back-end system development and has extensively worked with a wide range of database management systems

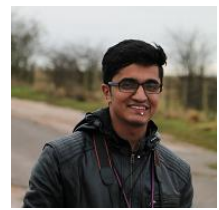

**Sudeepa Roy Dey** is a B.E and MTech in CSE and is currently pursuing Ph.D. in computer science and engineering from VTU.She is a currently working as Assistant professor in PESIT-BSC in CSE dept. Her research areas are **Machine** learning, Data Analytics, and Scientometrics.

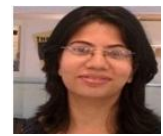

**Snehanshu Saha** holds Master's Degree in Mathematical Sciences at Clemson University and Ph.D. from the Department of Mathematics at the University of Texas at Arlington in 2008. After working briefly at his Alma matter, Snehanshu moved to the University of Texas El Paso as a regular full time faculty in the department of Mathematical Sciences. He is a Professor of Computer Science and Engineering at PESIT South since 2011 and heads the Center for Applied Mathematical Modeling and Simulation. He is also a visiting Professor at the department of Statistics, University of Georgia, USA. He has published 40 peer-reviewed articles in International journals and conferences and been IEEE Senior member, Governing Council member-International Astrostatistics Association and ACM professional member since 2012. Snehanshu's current and future research interests lie in Data Science and Machine Learning.

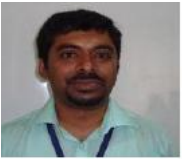

**Ankit Sinha** is currently pursuing a B.E. in Computer Science and Engineering from PESIT-BSC and is currently in his 3rd year. He is a part of Development team in the scibase project

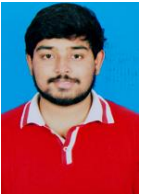